\documentclass[prl,showpacs,twocolumn,aps]{revtex4}
\usepackage{graphicx}
\usepackage{epsfig}
\usepackage[english]{babel}
\usepackage{amsmath}
\usepackage{amsfonts}
\usepackage{amssymb}%
\begin{document}
\title{Scaling of the von Neumann entropy across a finite temperature phase transition}
\author{Vladislav Popkov}
\author{Mario Salerno}
\affiliation{Dipartimento di Fisica "E.R. Caianiello" and
Consorzio Nazionale Interuniversitario per le Scienze Fisiche
della Materia (CNISM), Universit\'{a} di Salerno, I-84081,
Baronissi (SA), Italy}

\begin{abstract}
The spectrum of the reduced density matrix and the temperature
dependence of the von Neumann entropy (VNE) are analytically
obtained for a system of hard core bosons on a complete graph
which exhibits a phase transition to a Bose-Einstein condensate at
$T=T_c$.
It is demonstrated that the VNE undergoes a crossover from purely
logarithmic at $T=0$ to purely linear in block size $n$ behaviour
for $T\geq T_{c}$. For intermediate temperatures, VNE is a sum of
two contributions which are identified as the classical (Gibbs)
and the quantum (due to entanglement) parts of the von Neumann
entropy.
\end{abstract}
\pacs{03.67.Mn, 03.75.Gg, 87.15.Zg} \maketitle

\textit{ Introduction.} Recent experimental progresses in quantum
communication have increased interest on entanglement properties
of quantum systems both as resource for quantum computing and as
intrinsic characterization of quantum states \cite{Nielsen,
casati}. A key concept to quantify the entanglement of a part of a
system (e.g. a system of qubits) with respect to the rest
(environment) is the von Neumann entropy (VNE) which has been
shown to be related to the maximal compression rate of a quantum
information in an ideal coding scheme for two subsystems in a
joint pure state \cite{casati}. In the case of mixed states the
VNE contains contributions which come both from classical and
quantum (entanglement) correlations. In order to extract the
entanglement part from the VNE the classical contribution must be
subtracted, this requiring additional minimization over all
possible disentangled states (see, e.g.,
\cite{Vedral1997,VedralRMP}). The computational time needed to
perform such a minimization grows exponentially with the size of
the system, making the calculations impossible for any system with
more than a few particles.

In exceptional cases, however, the von Neumann entropy can be
computed exactly due to symmetries of the underlying quantum
problem. An example of this is given by integrable critical spin
chains which have ground states exactly known in the thermodynamic
limit, with long-distance correlations governed by conformal field
theory \cite{kitaev}. In this case VNE has been proved
\cite{Korepin,Cardy} to behave universally as $S_{(n)}=\frac{c}{3}
\log(\frac{1}{\pi T}\sinh(\pi n T )) +const$, in the limit of a
large block size $n$ and as a function of the temperature $T$,
where $c$ is the central charge of the underlying conformal field
theory.

In the present letter, we calculate, both analytically and
numerically,  the dependence on the temperature of the von Neumann
entropy for a class of quantum models which are invariant under
the action of the symmetric group and which arise in several
physical contexts, see e.g.
\cite{LMG,Toth,Penrose,Albertini,Mario,LMG05,Entanglement_higherspins}.
A common physical property of these models is that they exhibit a
phase transition from an ordered to a disordered state at a finite
critical temperature $T_{c}$ \cite{Penrose}. The knowledge of the
exact analytical expression of the VNE together with the
requirement that the classical part of the entropy must behave
extensively for large $n$, allow us to single out the entanglement
part of the VNE for arbitrary $n$ and $T$. As a result, we show
that the extensive part of the VNE coincides with the Gibbs
entropy computed directly from the energy spectrum, while the
entaglement part scales as $S_{(n)}-S_{Gibbs}=\frac{1}{2}\log n+
const$ below $T_{c}$ and as $S_{(n)}=S_{Gibbs}$ above $T_{c}$. For
$T=0$ the extensive part $S_{Gibbs}$ disappears and the result in
Ref. \cite{Entanglement_Heisenberg} is recovered.

\textit{The model.} As physical model we consider a system of hard-core bosons
described by the Hamiltonian
\begin{equation}
H=-\frac{1}{L}\sum_{i,j}^{L}b_{i}^{+}b_{j}+\sum_{i=1}^{L}b_{i}^{+}b_{i}\;,
\label{hardcore_bosons}%
\end{equation}
where $b_{i}^{\dagger},b_{i},$ denote creation and annihilation
operators satisfying
the following hard-core Heisenberg algebra: $[b_{i},b_{j}]=[b_{i}^{+},b_{j}^{+}%
]=0,[b_{i},b_{j}^{+}]=(1-2b_{j}^{+}b_{i})\delta_{ij}$. Due to the
on-site Fermi-like commutation relations, double occupancy is not
allowed: the action of $b_{i}^{+}$ and $b_{i}$ on the single
particle Fock space being
$b_{i}^{+}|0\rangle=|1\rangle;$ \ $b_{i}|1\rangle=|0\rangle;$ \ $b_{i}%
|0\rangle=b_{i}^{+}|1\rangle=0$. Note that the Hamiltonian
(\ref{hardcore_bosons}) is invariant under the action of the
permutation group (symmetric group $S_{L}$) and conserves the
number of particles $N=\sum _{i=1}^{L}b_{i}^{+}b_{i}$ with $N\leq
L$. The ground state of the system in the sector with $N$
particles is given by the  symmetric state
$|\Psi(L,N)\rangle=\binom{L}{N}^{-1/2}\sum_{P}|1111111...000000\rangle$
where sum is taken over all possible distributions of $N$
particles among $L$ sites.  Note that symmetric states are
intensively studied with symmetry techniques for testing  various
entanglement measures, see e.g. \cite{Hayashi2008} and references
therein. The excited states can be constructed from the
irreducible representations (irreps) of $S_{L}$ using filled Young
tableau (YT) of type $\{L-r,r\}$ with $r$ assuming all values in
the interval $[L/2]\geq r\geq1$ where $[x]$ denotes the integer
part of $x$ (see \cite{Mario,SE94}). The spectral and
thermodynamical properties of the system were studied in
\cite{Mario} where it was shown that eigenvalues of
(\ref{hardcore_bosons}) associated to YTs of type $\{L-r,r\}$ are
given by
\begin{equation}
E_{r}=r+\frac {1}{L}\left({N(N-1)-r(r-1)}\right),\;\;\; r=0,1,...,N\;, \label{eig}%
\end{equation}
with degeneracy $d_{r}=\binom{L}{r}-\binom{L}{r-1}$ \cite{Mario}.
An interesting property of the model is that it exhibits a phase
transition at finite temperature to a Bose-Einstein condensate
(BEC) of hard core bosons \cite{Penrose,Mario}. This can be
inferred directly from the free energy per site
$F/L=\Lambda_{\min}/\beta$ where
\begin{align}
\Lambda_{\min}  &  =\beta p^{2}+\min_{\mu \in
[0,\min(p,q)]}(\beta\mu(1-\mu)+\mu
\log\mu+\nonumber\\
&  (1-\mu)\log(1-\mu))\text{,} \label{free_energy}%
\end{align}
up to corrections of the order $o(L^{-1})$. Here and below we
denote $\beta=1/T$, $\mu=r/L$, $p=N/L$ and $q=1-p$. The extremum
condition for $\Lambda_{\min}$ leads to the equation
\begin{equation}
\beta^{\ast}(\mu^{\ast})=\frac{1}{(1-2\mu^{\ast})}\ln\left(
\frac{1-\mu ^{\ast}}{\mu^{\ast}}\right)\; ,
\label{transcendental}
\end{equation}
from which we see that for $T>T_{c}=(\beta^{\ast}(p))^{-1}$ there
is no solution for $\mu^{\ast}$ and the minimum of
(\ref{free_energy}) is reached at the end of the interval
$\mu=\min(p,q)$. For $T<T_{c}$, the minimum  of
(\ref{free_energy}) is inside the interval and is given by the
solution of (\ref{transcendental}). It was shown \cite{Penrose}
that this phase transition is actually a Bose-Einstein
condensation with the density of particles in the condensate given
by
\begin{equation}
\rho_{c}=(p-p^{\ast}(\beta))(q-p^{\ast}(\beta)) \label{density_condensate}%
\end{equation}
for $\ T<T_{c}$ and zero otherwise. In the following we shall characterize the
behavior of the VNE across this phase transition.

\noindent \textit{2. Behavior of the VNE across a classical phase
transition}. At T=0 the entanglement properties of a subsystem of
size $n$ with respect to the rest of the system (seen as
environment) can be characterized by the VNE
\begin{equation}
S_{(n)}=-tr(\rho_{(n)}\log_{2}\rho_{(n)})=-\sum\lambda_{k}\log_{2}\lambda_{k},
\label{von Neumann_entropy}%
\end{equation}
where $\lambda_k$ are the eigenvalues of the reduced density
matrix $\rho_{(n)}$, obtained from the density matrix $\rho$ of
the whole system as $\rho_{(n)}=tr_{(L-n)}\rho$. Since we are
interested in the behavior of the VNE across the finite
temperature BEC phase transition described above, we introduce the
thermal  von Neumann entropy for a block of size $n$ as follows
\begin{equation}
\label{S_n_finite_size} S_{(n)}(\beta)=\frac{1}{Z}\sum_{r=0}^{N}
d_{r} e^{-\beta E_{r}}Tr( \rho_{(n)}(r)\log_{2}\rho_{(n)}(r)),
\end{equation}
where $Z$ denotes the partition function. Note that at zero
temperature the density matrix of the whole system is a projector
on the completely symmetric ground state, $\rho=|\Psi(L,N)\rangle
\langle\Psi(L,N)|$ and due to the permutational symmetry,
$S_{(n)}$ does not depend on the choice of the sites in the block
but only on its size $n$. The corresponding VNE was obtained in
\cite{Entanglement_Heisenberg} where it was shown that
$\lambda_{k}=(^{n}_{k})(^{L-n}_{N-k})/(^{L}_{N})$, where
$k=0,1,...\min(n,N)$. In the limit of large $n$ von Neumann
entropy becomes
\begin{equation}
S_{(n)}\approx\frac{1}{2}\log_{2}(2\pi epq)+\frac{1}{2}\log_{2}%
\frac{n(L-n)}{L}. \label{entropy_L}
\end{equation}
To generalize (\ref{entropy_L}) to arbitrary temperatures we
remark that the reduced density matrix for temperatures $T>T_{c}$,
$L\rightarrow \infty$ is
\begin{equation}
\rho_{(n)}=\frac{1}{Z}\sum_{r=0}^{N}d_{r}e^{-\beta
E_{r}}\rho_{(n)} (r) \approx \rho_{(n)}|_{\mu=\min(p,q)}\;,
\end{equation}
while for $T<T_{c}$ we have $\rho_{(n)}=\rho_{(n)}(\mu^{\ast})$
with $\mu^{\ast}$ given by (\ref{transcendental}). Thus, to
compute the temperature-dependent von Neumann entropy we need to
know the eigenvalues of the reduced density matrix $\rho_{(n)}$
for arbitrary YT states.  The following theorem provides the
result.

\noindent \textit{3. Eigenvalues of $\rho_{(n)}$ and general
properties.}

\textbf{Theorem.} \textit{The eigenvalues of the reduced density
matrix $\rho_{(n)}$ of the eigenstates of H with N particles
belonging to the irreps of $S_{L}$ characterized by YTs of type
$\{L-r,r\}$, with $r< min(N,[L/2])$ are}
\begin{align}
&  \lambda(L,N,n,r,k,s)=\frac{(_{N-k}^{L-n})}{(_{N}^{L})}\;\;\sum_{i=0}%
^{k-s}(_{\;\;\;i}^{k-s})(_{\;\;\;\;\;i}^{n-k-s})\times
\label{Thermic_Eigen_Value}\\
&  \sum_{j=0}^{k-i}(-1)^{j}\frac{(_{j}^{s})}{(_{j+i}^{L-N})(_{j+i}^{\;N})}%
\sum_{m=0}^{j+i}(-1)^{m}(_{j+i-m}^{L-N-r})(_{j+i-m}^{\;N-r})(_{m}
^{r})\nonumber
\end{align}
\textit{with $k,s,$ quantum numbers assuming the values
$k=0,1,...,n$, and $s=0,1,...,\min(k,n-k)$. The corresponding
degeneracies coincide with the dimension of a YTs of type
$\{n,s\}$, i.e.} $\deg\lambda(L,N,n,r,k,s)=\binom
{n}{s}-\binom{n}{s-1}$.

The theorem follows from the block diagonalization of the reduced
density matrix with respect to the number of particles and to the
irreps of $S_{n}$. More precisely, $\rho_{(n)}$ can be block
diagonalized with respect to the number of bosons $k$ appearing in
the block, this leading to $n+1$ diagonal blocks $B_{k}$,
$k=0,1,...,n$. Each $B_{k}$ can be further diagonalized with
respect of the irreps of $S_{n}$ which are compatible with that
value of $k$, this leading to $k+1$ blocks associated to the YT of
type $\{n-s,s\}$ with $s=0,...,min(k,n-k)$. Notice that the above
decomposition implies that the block $B_{k}$ has dimension
$\sum_{i=0}^{k} \left( (^{n}_{i}) - (^{n}_{i-1})\right)=
(^{n}_{k})$ and the dimension of the matrix $\rho_{(n)}$ is $\sum_{k=0}%
^{n}(^{n}_{k}) = 2^{n}$, as it should be. This also clarifies the
meaning of the quantum numbers $k,s,$ and explains the
degeneracies given in the theorem. A full proof of the theorem
will be given elsewhere.

Before using the theorem we shall give some general properties of
the eigenvalues (\ref{Thermic_Eigen_Value}) and consider some
limiting cases from which the correctness of the result
(\ref{Thermic_Eigen_Value}) can be inferred.

\textit{a)} One can check by direct inspection that, as
consequence of the above block diagonalization, the eigenvalues in
(\ref{Thermic_Eigen_Value}) satisfy the following remarkable sums
with respect to $s$ and $k$
\begin{align}
&  \sum_{s=0}^{k}\lambda_{ks}deg(\lambda_{ks})=\frac{(_{N-k}^{L-n})(_{k}^{n}%
)}{(_{N}^{L})},\label{sumS}\\
&  \sum_{k=s}^{n-s}\lambda_{ks}=\frac{(_{r-s}^{L-n})-(_{r+s-n-1}^{L-n})}%
{(_{r}^{L})-(_{r-1}^{L})}, \label{sumK}%
\end{align}
where hereafter we use $\lambda_{ks}$ as a shorthand notation for
$\lambda (L,N,n,r,k,s)$. One can easily verify that the above sums
both lead to the correct normalization of $\rho_{(n)}$:
$tr\rho_{(n)}=1$. Indeed, by using Eq.(\ref{sumS}), we have
\begin{equation}
tr\rho_{n}=\sum_{k=0}^{n}\sum_{s=0}^{min(k,n-k)}\lambda_{k,s}deg(\lambda
_{ks})=\sum_{k=0}^{n}\frac{(_{N-k}^{L-n})(_{k}^{n})}{(_{N}^{L})}=1.\nonumber
\end{equation}
A similar expression is obtained by interchanging the order of the
sums and using Eq.(\ref{sumK}).

\textit{b)} Case $r=0$. In this case  in the sum over $m$ in
(\ref{Thermic_Eigen_Value}) only the term $m=0$ survives and,
since $r=0$ implies $s=0$, also the sum over $j$ has only one term
$j=0$. The eigenvalues (\ref{Thermic_Eigen_Value}) then reduce to
\begin{equation}
\label{lambda_k0}
\lambda_{k0}=\frac{(_{N-k}^{L-n})}{(_{N}^{L})}\;\;\sum_{i=0}^{k}(_{\,i}%
^{k})(_{\;\;\;\;\;i}^{n-k})=\frac{(_{N-k}^{L-n})}{(_{N}^{L})}(_{k}^{n})\;
,
\end{equation}
reproducing the exact result obtained for the completely symmetric
ground states in \cite{Entanglement_Heisenberg}.

Although explicit (i.e. summed) expressions for the eigenvalues
can be derived also for some other particular case (to be reported
elsewhere), a summed expression of Eq. (\ref{Thermic_Eigen_Value})
seems to be unlikely in the general case.

In the following we will work in the thermodynamic limit
$L\rightarrow\infty$ for which the expression for eigenvalues
simplifies drastically. Recalling that
$\lim_{L\rightarrow\infty}(r/L)=\mu$, $\lim_{L\rightarrow\infty
}(N/L)=p$, we obtain
\begin{eqnarray}
\label{lim1} &&
\lim_{L\rightarrow\infty}\frac{(^{L-n}_{N-k})}{(^{L}_{N})}=p^{n-k}q^{k},
\\ && \lim_{L\rightarrow\infty}\sum_{m=0}^{j}(-1)^{m}
\frac{(^{N-r}_{j-m})(^{L-N-r}_{\;\;
j-m})(^{r}_{m})}{(^{N}_{j})(^{L-N}_{\;\; j}) }=\eta^{j},
\end{eqnarray}
where
\begin{equation}
\label{eta} \eta=\frac{(p-\mu)(q-\mu)}{pq}
\end{equation}
plays the role of the order parameter $0\leq \eta \leq 1$, with
$\eta=0$ for $T\geq T_c$ and $\eta>0$ for $T\leq T_c$.
\begin{figure}[ptb]
\centerline{
\includegraphics[width=4.3cm,height=4.3cm,angle=0,clip]{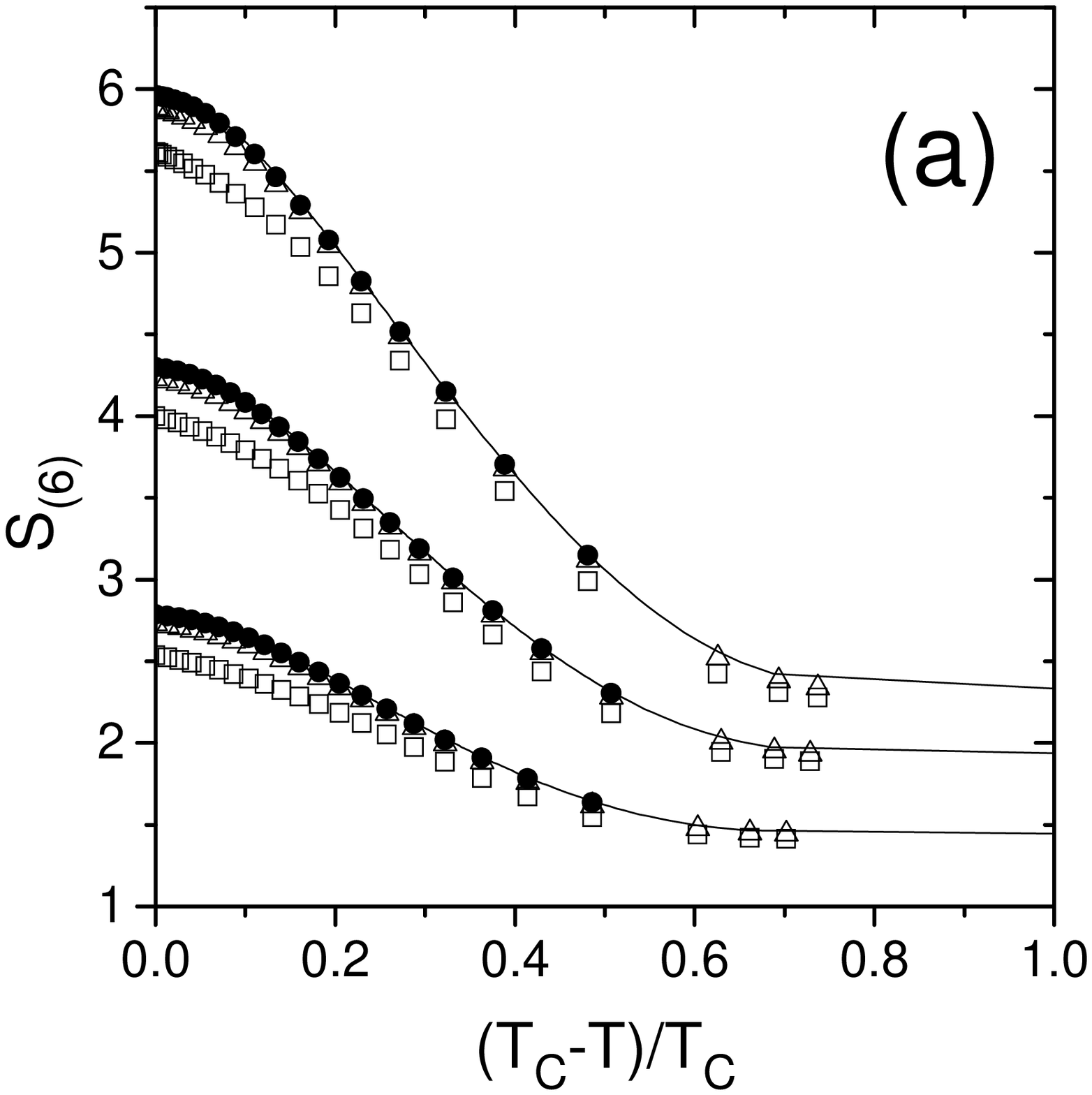}
\includegraphics[width=4.3cm,height=4.3cm,angle=0,clip]{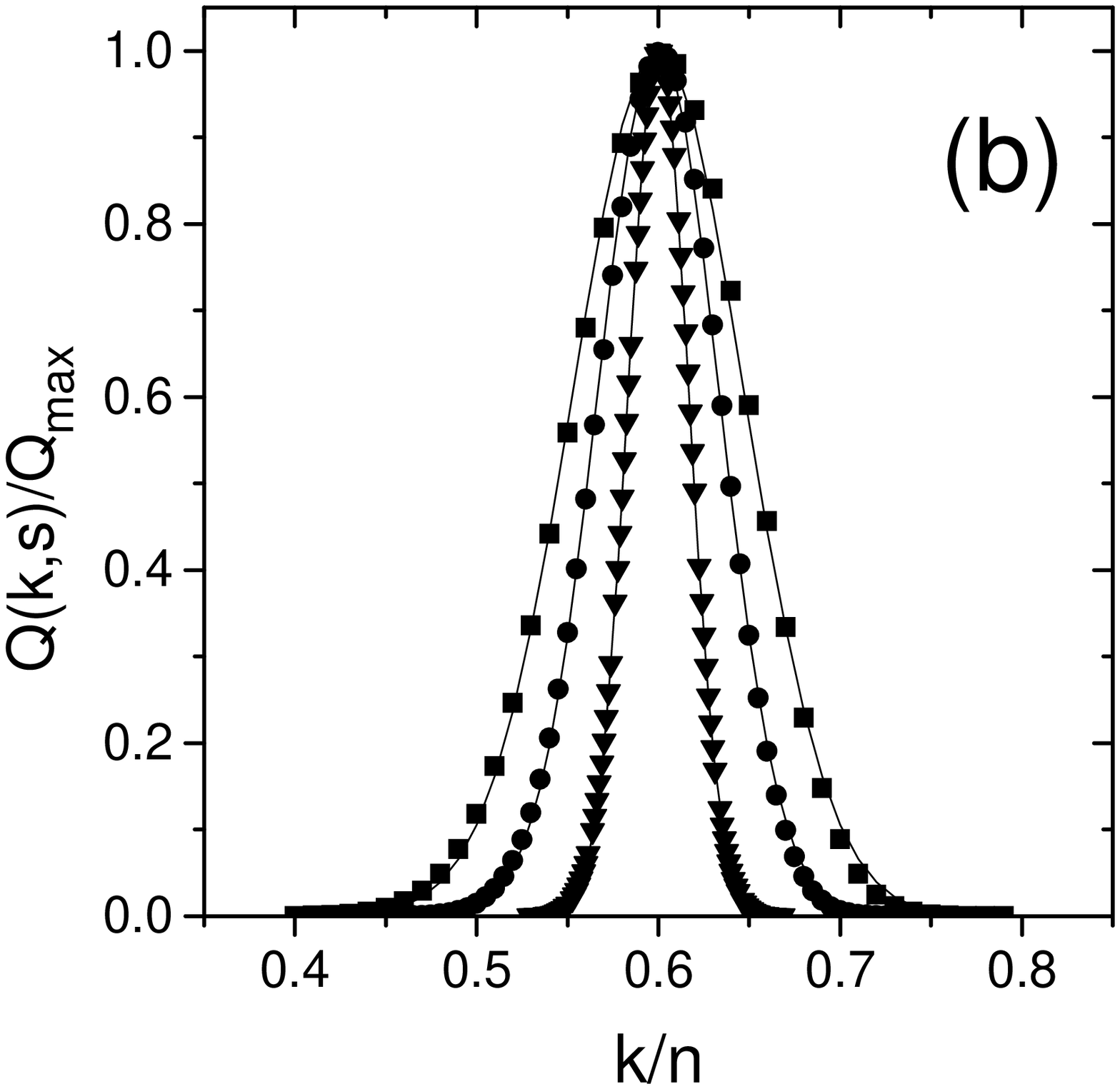}
}\caption{(a) Von Neumann entropy versus temperature for $n=6$,
obtained from Eq.(\ref{S_n_finite_size}) for different system
sizes: $L=50,200,700$ (squares, triangles and circles
respectively). Continuous curves denote  the corresponding
quantity in the thermodynamic limit
(\ref{eigenvalue_thermodynamic_hypergeometricsum}) while different
groups of curves correspond to different densities of hardcore
bosons in the system $N/L=0.1,0.2,0.5$ (from down up). (b)
Quantity $Q(k,s)/Q_{\max}=\lambda_{ks}
\deg(\lambda_{ks})/Q_{\max}$ computed from
(\ref{eigenvalue_thermodynamic_hypergeometricsum}) for fixed $s$,
versus $k/n$, for $n=100,200,800$ (squares, circles and triangles,
respectively). Parameters are $p=0.4,$ $\mu=0.2=s/n$. Curves are
given by the analytic prediction (\ref{Qks}).}
\label{Fig_S6thermicsquare}
\end{figure}
Inserting these expressions into (\ref{Thermic_Eigen_Value}) and
carrying out the summation
$\sum_{j=i}^{k}\eta^{j}(-1)^{j}(^{\;\;s}_{j-i})=(-\eta)^{i}(1-\eta)^{s}$
we obtain
\begin{align}
\lim_{L\rightarrow\infty}\lambda_{ks} =
p^{n-k}q^{k}(1-\eta)^{s}\sum
_{i=0}^{k-s}\eta^{i}(^{k-s}_{\;\;i})(^{n-k-s}_{\;\;\;i})=\nonumber
\\ = p^{n-k}q^{k}(1-\eta)^{s} \text{}_{2}F_{1}(-k+s,k-n+s;1;\eta),
\label{eigenvalue_thermodynamic_hypergeometricsum}%
\end{align}
where $_{2}F_{1}(a,b;c;d)$ is Gauss hypergeometric function. In
Fig.\ref{Fig_S6thermicsquare}(a) we compare the von Neumann
entropy in (\ref{S_n_finite_size}) with the thermodynamic limit
(\ref{eigenvalue_thermodynamic_hypergeometricsum}), for different
temperatures and various particle densities.

\noindent \textit{4. Classical and quantum parts of the von
Neumann entropy.} Various limits of the above formula are
discussed below.

\noindent i) Zero temperature limit $T\rightarrow0$. In this case
$\eta=1$, and from
(\ref{eigenvalue_thermodynamic_hypergeometricsum}) we have
$\lambda_{k0}=p^{n-k}q^{k}\binom{n}{k}$. This is just the limit
$L\rightarrow\infty$ of Eq. (\ref{lambda_k0}), reproducing the
results obtained in \cite{Entanglement_Heisenberg}.

\noindent ii) High temperatures $T\geq T_{c}$ . In this case
$\eta=0$ and the eigenvalues
(\ref{eigenvalue_thermodynamic_hypergeometricsum}) become
$s$-independent $\lambda_{ks}=p^{n-k}q^{k}$. This leads to the
extensive classical entropy (Gibbs entropy) of an ideal gas with
excluded volume
\begin{equation}
S_{(n)}(T\geq T_{c})=-n(p\log p+(1-p)\log(1-p)). \label{Tc_limit}
\end{equation}

\noindent iii) Intermediate temperatures $0 < T < T_{c}$. This
case $0<\eta<1$ is the most interesting one since the VNE has both
classical (due to degeneracy) and quantum (due to entanglement)
contributions. Analyzing the sum in
(\ref{eigenvalue_thermodynamic_hypergeometricsum}) one can show
that the variable
$Q(k,s)=\lim_{L\rightarrow\infty}\lambda_{ks}\deg(\lambda_{ks})$
is Gaussian-distributed with the mean $\langle k/n\rangle=q$,
$\langle s/n\rangle=\mu$:
\begin{equation}
Q(k,s)\approx Q_{max} e^{-[{\frac{(s-n\mu) ^{2}}{2C}+\frac{(k-nq)
^{2}}{2D}+\frac{( s-n\mu)(k-nq) }{B}}]}, \label{Qks}
\end{equation}
with $Q_{max}=(1-2\mu)(2\pi n\sqrt{\mu(1-\mu)(p-\mu)(q-\mu)})$,
and
\begin{eqnarray}
&&
C^{-1}=\frac{1}{n\mu(1-\mu)}+\frac{(p-q)^{2}}{n(p-\mu)(q-\mu)},\quad
\nonumber
\\
&& D^{-1}=\frac{(1-2\mu)^{2}}{n(p-\mu)(q-\mu)}, \;
B^{-1}=\frac{(1-2\mu )(p-q)}{n(p-\mu)(q-\mu)}. \nonumber
\end{eqnarray}
In Fig.\ref{Fig_S6thermicsquare}(b) we compare the distribution of
the eigenvalues obtained from direct calculations with the
Gaussian distribution (\ref{Qks}), from which we see that the
agreement is excellent. Using the above expression of $Q(k,s)$ we
calculate the VNE as
\begin{equation}
S_{(n)}=-{\displaystyle\sum\limits_{k=0}^{n}}{\displaystyle\sum\limits_{s=0}%
^{\min(k,n-k)}}Q(k,s)\log_{2}\frac{Q(k,s)}{\deg\lambda_{ks}}.
\label{Entropy_T}%
\end{equation}
\begin{figure}[ptb]
\centerline{
\includegraphics[width=4.3cm,height=4.3cm,angle=0,clip]{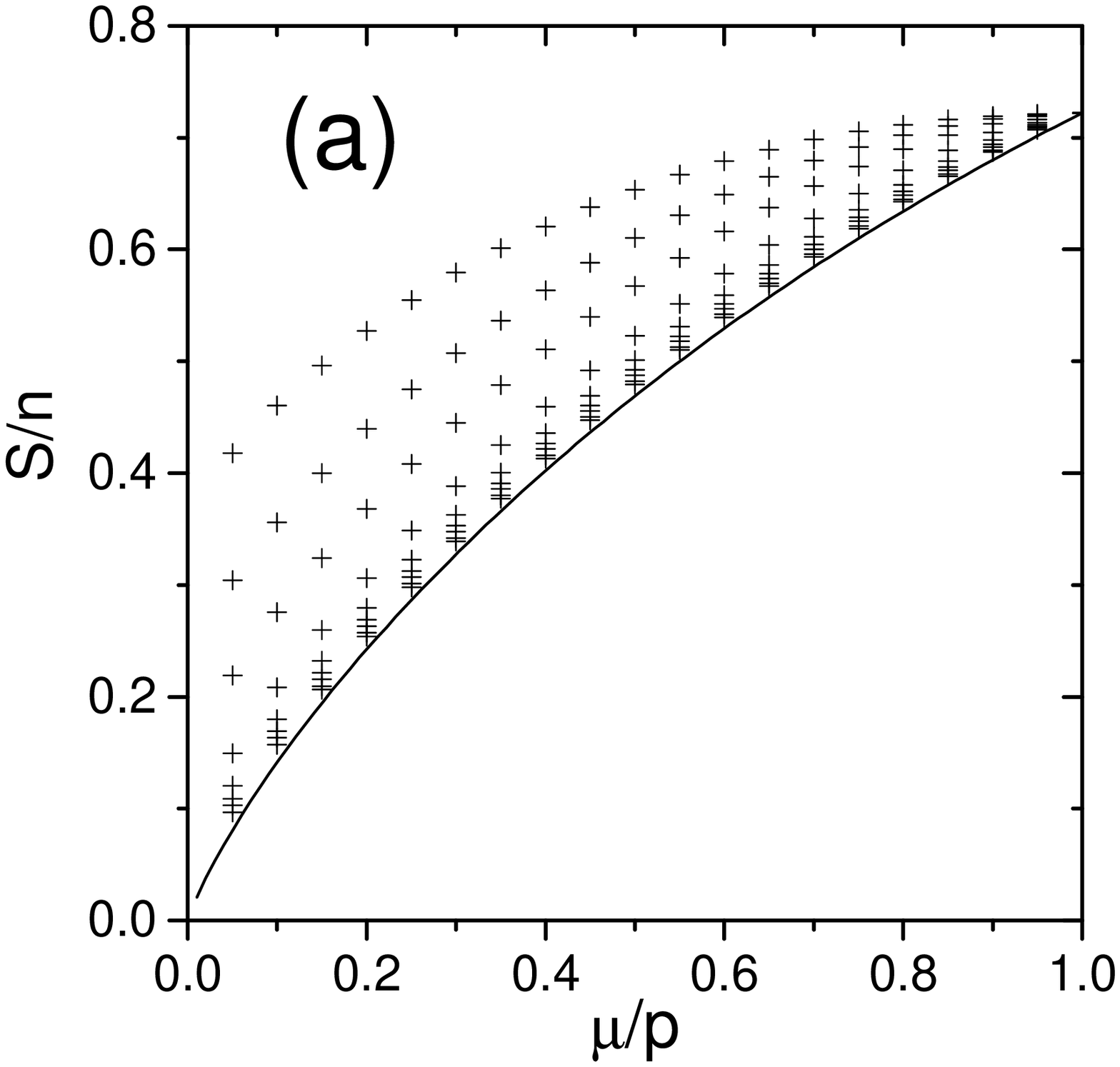}
\includegraphics[width=4.3cm,height=4.3cm,angle=0,clip]{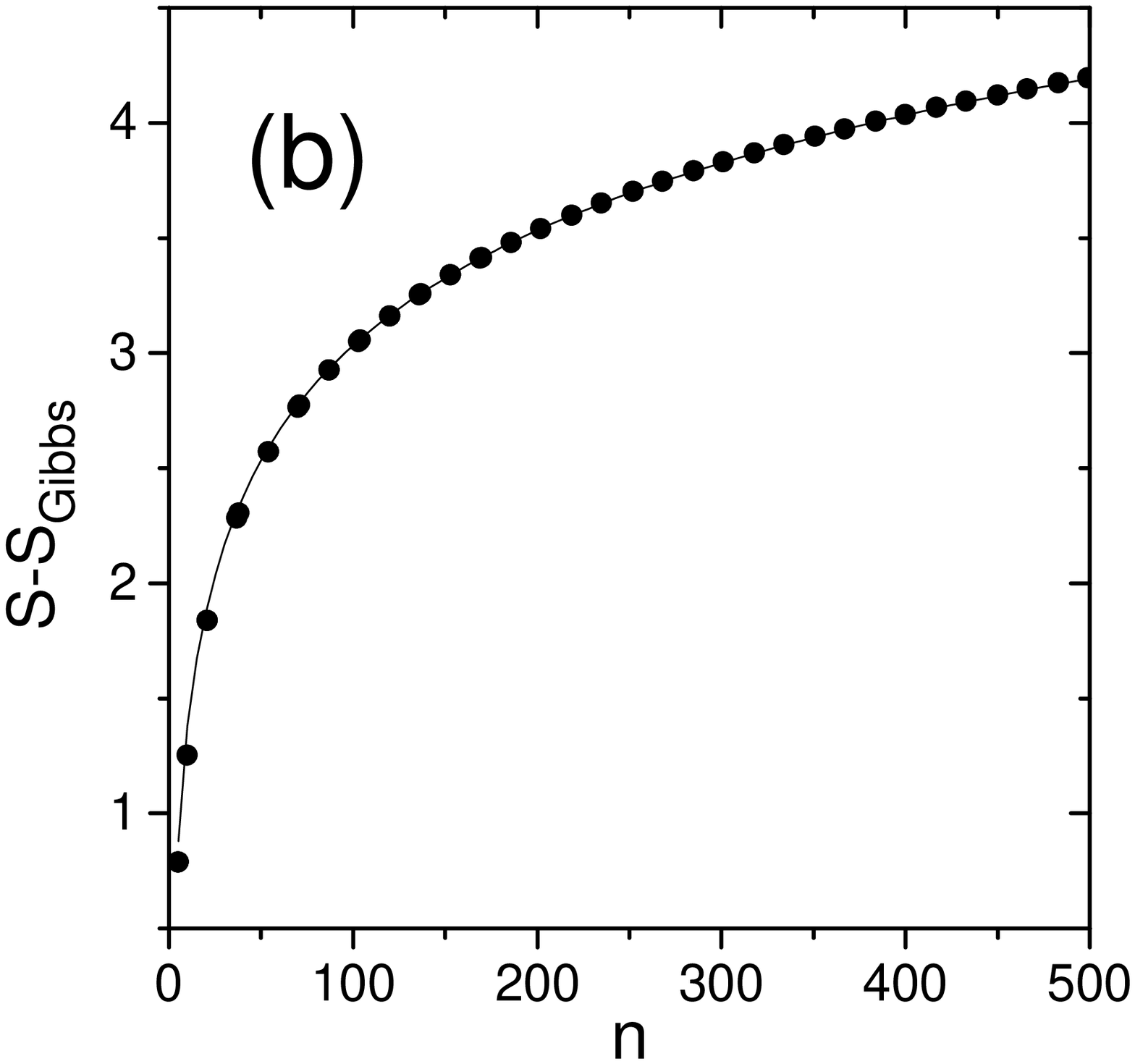}
}\caption{(a). Extensive part of the von- Neumann entropy of a
block $\lim_{n\rightarrow\infty}S_{(n)}/n$ versus renormalized
temperature $\mu^{\ast}/p$ for $p=0.2$. Comparison of analytic
prediction (\ref{Gibbs_entropy}) (curve) with the exact
calculations from
(\ref{eigenvalue_thermodynamic_hypergeometricsum}) for finite
$n=5,10,50,100,150,200,300,400$ (points approaching the curve from
above). (b). "Quantum" part of the von Neumann entropy
$S(n)-S_{Gibbs}(n)$ versus block-size $n$, for $p=0.3$ and
$\mu^{\ast}=0.18$ (points). The continuous curve refers to the
fitting
function $\frac{1}{2}\log_{2}n-0.289$.}%
\label{Fig_S6thermic}%
\end{figure}
Substituting the sums with the integrals and using the
normalization ${\displaystyle\sum}Q(k,s)  =1$ we obtain
\begin{equation}
S_{(n)}\approx-n(\mu\log_{2}\mu-(1-\mu) \log_{2}(1-\mu))
+\frac{1}{2}\log_{2}n+R(q,\mu)
\label{entropy_thermodynamic}%
\end{equation}
where $R(q,\mu)$ is a non-universal $n$-independent constant. The
first term at the right-hand side\ of
(\ref{entropy_thermodynamic}) gives the major contribution in the
large $n$ limit, and coincides with the Gibbs entropy obtained
from the spectrum of the whole system (\ref{eig}) in the
thermodynamic limit $E(L,\mu,p)/L\approx\mu-\mu^{2}+p^{2}$,
$d_{r}\approx\frac{1-2\mu}{(1-\mu)^{L-r+1}\mu^{r}}$, as
\begin{equation}
\frac {S_{Gibbs}}{L}=\lim_{L\rightarrow\infty} \frac{ \beta
\langle
E-F\rangle}{L}, \label{Gibbs_entropy_definition}%
\end{equation}
where $F/L=\Lambda_{\min}/\beta$ and $\Lambda_{\min}$ is given by
(\ref{free_energy}). Inserting the expressions for $E$ and $F$
into (\ref{Gibbs_entropy_definition}), we obtain
\begin{equation}
S_{Gibbs}/L=-\mu^{\ast}\log\mu^{\ast}-\left(  1-\mu^{\ast}\right)
\log (1-\mu^{\ast}) \label{Gibbs_entropy}
\end{equation}
where $\mu^{\ast}(T)$ is defined in Eq.(\ref{transcendental}). The
term $\frac{1}{2}\log_{2}n$ in (\ref{entropy_thermodynamic})
coincides, up to a constant, with the entanglement entropy at zero
temperature (\ref{entropy_L}) and can therefore be interpreted as
entanglement entropy part in the von Neumann entropy at finite
temperatures, defined as
\begin{equation}
\label{Sentang}
S_{\text{ent}}(n,T)=S_{(n)}-n\lim_{L\rightarrow\infty}\frac{E-F}{TL}=\frac
{1}{2}\log_{2}n+R(q,\mu).
\end{equation}
Note that this contribution disappears at $T_{c}$ and above. In
Fig.\ref{Fig_S6thermic} the extensive (panel (a)) and "quantum"
parts (panel(b)) of the VNE  are depicted.

\noindent \textit{5 Conclusions.} We have calculated the spectrum
of the reduced density matrix and von Neumann entropy of a block
of $n$ sites, for a quantum model with permutational symmetry, as
function of temperature and particle density. It is shown that
eigenvalues of the reduced density matrix in the thermodynamic
limit are parametrized by the single parameter defined in
(\ref{eta}), which also turns out to be the order parameter of the
problem. We defined the entropy of entanglement for finite
temperature as the difference between the VNE and its extensive
part (coinciding with the Gibbs entropy), and demonstrated its
disappearance above the classical phase transition. For all
temperatures below the critical one, the entropy of entanglement
scales as $\frac{1}{2}\log n$. We expect this results to be valid
also for other quantum systems with permutational symmetry (mean
field models) exhibiting finite temperature phase transitions.

\textit{Acknowledgements.} V.P. thanks the Department of Physics
of the University of Salerno for a research grant (Assegno di
Ricerca n.1013-2006) during which this work was done.

\end{document}